\documentclass[a4paper,11pt]{article}
\pdfoutput=1 

\usepackage{jheppub} 

\usepackage[T1]{fontenc} 

\usepackage{amssymb}
\usepackage{graphicx}
\usepackage{amsmath,epsfig}
\def\be{\begin{equation}}
\def\ee{\end{equation}}
\def\bea{\begin{aligned}}
\def\eea{\end{aligned}}
\def\ba{\begin{eqnarray}}
\def\ea{\end{eqnarray}}

\usepackage[usenames]{color}

\def\yzero{\smash{\hbox{$y\kern-4pt\raise1pt\hbox{${}^\circ$}$}}}

\def\-{\hphantom{-}}

\def\s2{\frac{1}{\sqrt2}}

\def\IF{\relax{\rm I\kern-.18em F}}
\def\II{\relax{\rm I\kern-.18em I}}
\def\IP{\relax{\rm I\kern-.18em P}}
\def\IC{\relax\hbox{\kern.25em$\inbar\kern-.3em{\rm C}$}}
\def\IR{\relax{\rm I\kern-.18em R}}

\def\Dsl{\,\raise.15ex\hbox{/}\mkern-13.5mu D} 
\def\IZ{Z\kern-.4em  Z}

\title{Fluxes, Twisted tori, Monodromy and $U(1)$ Supermembranes}
\author[a]{M.P. Garcia del Moral}
\author[a]{C. Las Heras}
\author[a]{P. Leon}
\author[a]{J.M. Pena}
\author[a]{A. Restuccia}

\affiliation[a]{Departamento de F\'isica, Universidad de Antofagasta, \\ Aptdo 02800, Chile.}

\emailAdd{maria.garciadelmoral@uantof.cl}
\emailAdd{camilo.lasheras@ua.cl}
\emailAdd{pablo.leon@ua.cl}
\emailAdd{joselen@yahoo.com}
\emailAdd{alvaro.restuccia@uantof.cl}

\abstract{
We show that the $D=11$ Supermembrane theory (M2-brane) compactified on a $M_9 \times T^2$  target space, with constant fluxes $C_{\pm}$ naturally incorporates the geometrical structure of a twisted torus. 
We extend the M2-brane theory to a formulation on a twisted torus bundle. It is consistently fibered over the world volume  of the M2-brane. It can also be interpreted as a torus bundle with a nontrivial $U(1)$ connection associated to the fluxes. The structure group $G$ is the area preserving diffeomorphisms. The torus bundle is defined in terms of the monodromy  associated to the isotopy classes of symplectomorphisms  with $\pi_{0} (G) = SL(2,Z)$, and classified by the coinvariants of the subgroups of $SL(2,Z)$.  The spectrum of the theory is purely discrete since the constant flux induces a central charge on the supersymmetric algebra and a modification on the Hamiltonian which renders the spectrum discrete with finite multiplicity. The theory is invariant under symplectomorphisms connected and non connected to the identity, a result relevant to guaranteed the U-dual invariance of the theory.  The Hamiltonian of the theory exhibits  interesting new $U(1)$ gauge and global symmetries on the worldvolume induced by the symplectomorphim transformations. We construct explicitly the supersymmetric algebra with nontrivial central charges.  We show that the zero modes decouple from the nonzero ones. The nonzero mode algebra corresponds to a massive superalgebra that preserves either $1/2$ or $1/4$ of the original supersymmetry depending on the state considered.}

\begin{document} 
\maketitle
\flushbottom

\section{Introduction} 
Flux compactifications has provided a new arena to explore String theory realizations with very good results towards the recovering of phenomenological properties at low energies. They are topological quantities associated to the quantization condition of closed field strength p-forms over compact p-cycles. They may modify the string theory compactifications in many ways, for example in the amount of supersymmetry preserved, by providing a mechanism for perturbative moduli stabilization, generating susy/soft breaking terms, a hierarchy of scales between gravitational and gauge forces, as well as deforming the geometry associated to the compactified manifold.  It has also been showed that string compactifications with fluxes may generate gauged supergravities as Effective Field Theory (EFT). 

String/ F-theory compactifications on twisted torus have been proved to provide very fruitful scenarios to obtain at low energies gauged supergravities \cite{Hull4, DallAgata, DallAgata2, Trigiante, Kaloper, Hull5, ReidEdwards, Grana, Shelton, DAuria} since its proposal, in  \cite{Schwarz5}, as a mechanism for  nontrivial reducing 11D supergravity to lower dimensions. In those works it was established the relation between  $11D$ Supergravity reductions, by Sherk-Schwarz reductions \cite{Schwarz5, Bergshoeff5, Lowe, ReidEdwards2}, and geometric fluxes, also call torsion  \cite{Andrianopoli}, or by more general fluxes described  via embedding tensor mechanism \cite{Samtleben,  Melgarejo}. String compactifications on twisted tori can be described in two complementary ways \cite{Jonke}: as a group manifold \cite{Hull6, Hull9} (a nilmanifold \cite{Jonke, Thangavelu,Shi}) or as T-duals of tori with constant NS 3-form flux \cite{Kachru}. The twisted torus seen as fiber bundles is associated to nontrivial torus bundles with monodromy over torus. In the particular case of twisted $T^3$ it can be described either as nontrivial $U(1)$ principal bundles over a torus \cite{Pope, Shape} or as a 2-torus bundle over a circle with monodromy in $SL(2,Z)$  \cite{Kaloper}. 

Recently in the context of the Supermembrane theory,ie. the  M2-brane theory, as part of M-theory, the relevance  of flux compactification in the quantum behaviour of the theory has been emphasize \cite{mpgm6}. However other implications have not been reported. A way to obtain nine dimensional type II gauged supergravities as an Effective Field Theory (EFT) at low energies has been from F-theory compactified on a 3-twisted torus (\cite{Hull9}) or from M-theory \cite{Hull8}. In this paper we will show that the supermembrane with constant fluxes $C_{\pm}$ compactified on $M_9\times T^2$ can be understood as a Supermembrane compactificatified on a twisted tori, providing  new hints of its connection \cite{mpgm2},\cite{mpgm7} with type II gauged supergravities in nine dimensions \cite{Bergshoeff5},\cite{Melgarejo}. 

The M2-brane formulated on a twisted torus bundle considers the twisted torus $\mathcal{T}_W^3$ to contain two of it dimensions associated to the 2-torus target space and the third one to the fiber of the nontrivial $U(1)$ principal bundle associated to a central charge on the worldvolume.
Geometrically it can be considered as an emergent internal dimension that will play no dynamical role. The $U(1)$ principal bundle associated to the nontrivial flux on the target can be pullback by an immersion map defined in terms of the harmonic 1-forms on the base torus $\Sigma$. It turns out that this map is a diffeomorphism between the torus $\Sigma$ and the flat torus $T^2$ on the target space. Hence the $U(1)$ principal bundle on $T^2$ can be pullback to a $U(1)$ principal bundle on $\Sigma$ and viceversa. 

This bundle has a $U(1)$ connection 1-form on $\Sigma$ whose curvature has an associated Chern number characterizing a central charge on the supersymmetric algebra of the Supermembrane. We discuss the explicit relation between the geometry associated to the flux condition on the target and the central charge on the base torus $\Sigma$. It was shown in \cite{mpgm6} the equivalence between both geometrical structures. 

Moreover, we complement this result with the construction of the supersymmetric algebra of the Supermembrane with constant fluxes $C_{\pm}$. We make explicit the supersymmetric algebra following the lines of  \cite{dwhn}. We discuss the decoupling of the zero modes for this case and end up with in the non zero mode  algebra with the Hamiltonian of the Supermembrane with central charges. 

We discuss the amount of supersymmetry preserved by the flux or central charge condition. Besides the $U(1)$ geometrical structure which we explicitly discuss, the Supermembrane theory is invariant under area preserving diffeomorphisms APD  connected to the identity. In fact, the Hamiltonian of the Supermembrane is subject to a first class constraint which generates that symmetry of the theory. 

In two dimensions the area preserving diffeomorphism group coincides with the symplectomorphism group preserving a symplectic 2-form, which in our construction of the Supermembrane with central charges is the induced symplectic 2-form arising from the pullback of the canonical symplectic 2-form on the target torus $T^2$ by the diffeomorphism between the two tori.

In the presence of a central charge on the worldvolume, the formulation of the Supermembrane introduces an explicit dependence on the homology basis on $\Sigma$ as well as on the harmonic 1-form basis on $\Sigma$ and on the three parameters characterizing $T^2$, in particular the Teichmuller $\tau$ associated to the period of the normalized holomorphic 1-form.

It is then relevant to analyse the dependence of the Supermembrane theory on the non-connected to the identity symplectomorphisms. 

We will show that, the hamiltonian of the M2-brane with fluxes is invariant under the full group of symplectomorphisms and under $U(1)$ gauge symmetry associated to the fluxes or equivalently to the central charge on the worldvolume. The Hamiltonian can be expressed either in terms of a symplectic connection and its curvature or in terms of a $U(1)$ connection and its curvature, a remarkable property. All these aspects can be of relevant for realistic compactifications.

Moreover, we will discuss the compatibility between both geometrical structures the symplectic one and the $U(1)$ principal bundle and the relation of these geometrical structures with the formulation of the Supermembrane on a twisted torus bundle with monodromy generated by a representation of the fundamental group of $\Sigma$ acting on $H_1(T^2)$, the first homology group of $T^2$, which can be identified with the natural action of $SL(2,Z)$ on $\mathcal{Z}^2$.

The main point to be shown is the consistency of the transitions on the symplectic group under the monodromy and the transitions on the $U(1)$ line bundle. The formulation of the Supermembrane in terms of the geometrical objects of a twisted torus bundle will then follow directly. If that is so, the classification of inequivalent classes of M2-brane torus bundles already known \cite{mpgm2,mpgm7} and related to gauge supergravities in \cite{Bergshoeff5,Melgarejo} can be enriched by the associated monopole structure \cite{Restuccia2} which can be relevant for applications of the M2-brane phenomenology.

Finally, a remarkable property of M2-branes on a twisted torus bundle is that the quantum spectrum of their regularized Hamiltonian is discrete. This follows from the presence of a nontrivial central charge condition \cite{Restuccia}  that render the spectrum purely discrete with finite multiplicity \cite{Boulton}. The same results occurs for  the M2-brane with nontrivial constant fluxes $C_{\pm}$ on the target \cite{mpgm6}. 

This paper is organized as follows. In section 2. we  summarize the formulation of the M2-brane on a $M_9\times T^2$ target space in the presence of constant  fluxes $C_{\pm}$ and its relation with the  so-called central charge condition.
In section 3.  we obtain the algebra of supercharges of the supermembrane with fluxes and analyze the behaviour of zero and non-zero modes.We also discuss the amount of supersymmetry preserved by the nonzero algebra of supercharges. In section 4 we discuss the symplectomorphims transformations connected and non connected to the identity under which the Hamiltonian is invariant. In section 5. we obtain the gauge and global $U(1)$ symmetries of the Hamiltonian associated to symplectomorphism transformations. We also obtain a symplectic gauge symmetry realized on the Hamiltonian. We obtain two different formulations of the Hamiltonian depending on which gauge symmetry we want to make manifest.  In section 6.  we obtain a geometrical interpretation of the M2-brane formulated on a twisted torus bundle. We discuss its consistency. In section 7 we present a discussion and our conclusions.


\section{ M2-brane with constant  $C_{\pm}$ fluxes}
In this section we review former results found in \cite{mpgm6} where it was shown that a M2-brane with $C_{\pm}$ fluxes  can be interpreted as a M2-brane in a background with central charges. 
 The supersymmetric action of the M2-brane on a generic 11D noncompact background was found by \cite{Bergshoeff}. 
In the following we will consider a flat background metric $G_{\mu \nu} =\eta_{\mu \nu}$  in the presence of some constant components of the three-form $C_3$.
The embedding coordinates in the superspace formalism are $(X^{\mu}(\xi),\theta^{\alpha}(\xi))$ with $\xi^r $ the worldvolume coordinates and where $\mu, \nu, \lambda $ and  $\alpha, \beta,$ are bosonic and fermionic target space indices, respectively and $r,s,t$ denote worldvolume indices.
In this background the action of the supermembrane takes the following form
\begin{equation}
\small
\begin{aligned}
\label{ec1}
S= & - T \int d^3 \xi \{ \sqrt{-g}+\varepsilon^{rst}\bar{\theta}\Gamma_{\mu \nu}\partial_t \theta \left[ \frac{1}{2}\partial_r X^\mu (\partial_s X^\nu\right. +  \bar{\theta}\Gamma^\nu \partial_s\theta) + \\
& + \frac{1}{6}\bar{\theta}\Gamma^\mu \partial_r \theta \bar{\theta}\Gamma^\nu \partial_s \theta ] +\frac{1}{6}\varepsilon^{rst}\partial_r X^\mu \partial_s X^{\nu} \partial_t X^\rho C_{\rho \nu \mu} \} \, ,
\end{aligned}
\end{equation}
This background corresponds to the asymptotic limit of a $D=11$ supergravity solution generated by an M2-brane acting as a source \cite{Duff4, Stelle}. 
We notice that the last term in (\ref{ec1}) becomes non trivial because the maps $X^\mu$ may have a nontrivial wrapping on the compact torus of the target space.

We now consider the supermembrane action in the Light Cone Gauge (LCG) on a $M_{11}$ target space with constant gauge field $C_{{{\mu}}{{\nu}}{{\lambda}}} $ closely following the definitions in \cite{deWit}. The supersymmetric action is \cite{mpgm6},
\begin{equation}
\small
S = T\int d^3 \xi \{ -  \sqrt{\bar{g}\Delta}-\varepsilon^{uv}\partial_uX^a \bar{\theta} \Gamma^- \Gamma_a \partial_v \theta +  C_+ +\partial_{\tau} X^- C_- +\partial_{\tau} X^a C_a+C_{+-} \}
\end{equation}
with
\begin{equation}
\small
\begin{aligned}
& C_a  =  -\varepsilon^{uv}\partial_uX^- \partial_vX^b C_{-ab} +\frac{1}{2}\varepsilon^{uv}\partial_uX^b \partial_vX^c C_{abc} \, , \\
& C_{\pm}  =  \frac{1}{2}\varepsilon^{uv}\partial_uX^a \partial_vX^b C_{\pm ab} \,, \qquad C_{+-}  =  \varepsilon^{uv}\partial_uX^- \partial_vX^a C_{+-a} \,,
\end{aligned}
\end{equation}
$a,b,c=1,...,9$ are target space transverse coordinates indices, and $u,v=1,2$ are indices for worldvolume spacelike coordinates $(\sigma^1,\sigma^2)$. It is possible to fix the variation of some components of the 3-form by virtue of its gauge invariance. In particular it is possible to fix $C_{+-a}=0$ and $C_{- ab}=constant.$  
%
The action contains nonphysical degrees of freedom $X^{-}$ that must be eliminated. In \cite{mpgm6} the dependence on $X^-$ was eliminated by performing a canonical transformation on the configuration variables. On the new variables the Hamiltonian of the compactified theory on $M_9 \times T^2$ target space is the following one:
\begin{equation}
\small\small
\label{hamiltonian-fluxes}
\begin{aligned} 
\widetilde{H} = T\int_{\Sigma} d^2\sigma \{ \frac{\sqrt{w}}{\hat{P}_{-}^0} \, \left[ \frac{1}{2} \left(\frac{P_m}{\sqrt{w}} \right)^2 + \frac{1}{2}\left(\frac{P_i}{\sqrt{w}} \right)^2 + \frac{1}{4} \left\{ X^i, X^j \right\} ^2 +  \frac{1}{2} \left\{ X^i, X^m \right\}^2  \right. \\
\left. + \frac{1}{4} \left\{ X^m, X^n \right\} ^2 \right] +\sqrt{w} \left[ \bar{\theta}\Gamma^-\Gamma_m\left\lbrace X^m,\theta\right\rbrace +\bar{\theta}\Gamma^-\Gamma_i\left\lbrace X^i,\theta\right\rbrace\right]  -C_+ \} \,,
\end{aligned}
\end{equation}
where the  $X^m, m=3,\dots,9$ denote the transverse maps from the foliated worldvolume $\Sigma$ to $M_9$ and $X^i, i,j=1,2$ the maps  from $\Sigma$ to $T^2$ and the Lie bracket is defined as  $\{A,B\}=\frac{\epsilon_{uv}}{\sqrt{w}}\partial_uA\partial_v B$. In the compactified case, in contrast to the noncompact one, the last term in (\ref{hamiltonian-fluxes}) for constant bosonic 3-form is a total derivative of a multivalued function, therefore its integral is not necessarily zero.
This Hamiltonian (\ref{hamiltonian-fluxes}) is subject to the local and global constraints associated to the area preserving diffeomorphisms (APD) connected to the identity 
\begin{equation}
\small
\label{constraintdos}
d(P_i dX^i + P_m dX^m+\overline{\theta}\Gamma^{-}d\theta)=0 \,, \qquad
 \oint_{{\mathcal{C}}_s} (P_i dX^i + P_m dX^m+\overline{\theta}\Gamma^{-}d\theta)=0 \,.
\end{equation}

Classically the dynamics of this Hamiltonian contains string-like spikes which render the quantum spectrum of the theory continuous.

%
%
When the following flux condition is imposed on  the target space $M_9 \times T^2$  associated to the nontrivial integral values of the two form $\widetilde{F} = \frac{1}{2}C_{\pm ij}d\widetilde{X}^i\wedge d\widetilde{X}^j$ 
\begin{equation}
\small
\label{flux condition}
\int_{T^2}{\widetilde{F}}
= k_{\pm} \in \mathbb{Z}/\{0\} \,,
\end{equation}
with $d\widetilde{X}^i, i=1,2$  the harmonic one-forms of the $T^2$ and we impose that the maps $X^i$ from the worldvolume to the compact sector of the target are identified with $\widetilde{X}^i$, then the spectrum of the Supermembrane  becomes discrete. This flux condition is equivalent to the existence of an $U(1)$ principle bundle over $T^2$ and of a 1-form connection on it, whose curvature is $\widehat{F}$.
In particular, the pullback of (\ref{flux condition}) by the  embedding maps describes the central charge condition over the supermembrane worldvolume. In \cite{mpgm6} we proved that : 
\begin{equation}
\small
\label{fluxcondition2}
\int_{T^2}\widetilde{F}=\int_{\Sigma}\widehat{F} = \int_{\Sigma} \frac{1}{2}\epsilon_{ij}dX^i \wedge dX^j = n ,\quad n\in \mathbb{Z}/\{0\} \,,
\end{equation}
where $X^i$ are the maps from $\Sigma$ to the $T^2$ on the target, and for $k_{\pm}=n$ (and $C_{\pm ij}=\epsilon_{ij}$). Then, there is a one to one correspondence between the supermembrane where $\widehat{F}$ is the curvature on the world volume associated to irreducible winding of the membrane, and the supermembrane on a background with a flux condition on $T^2$ generated by ${C_{\pm}}$.
The irreducible wrapping condition, when the area of the $T^2$ has been normalized to one, is 
\begin{equation}
\label{cc}
    \int_{\Sigma} dX^i\wedge dX^j = \epsilon^{ij}n, \quad n\in \mathbb{Z}/\{0\} \,,
\end{equation}
is a nontrivial 2-form flux condition over the worldvolume that generalizes the Dirac monopole construction to Riemann surfaces of arbitrary genus $g\ge 1$ \cite{Restuccia2} applied to the supermembrane theory \cite{Restuccia}. The invariance under area preserving diffeomorphisms is preserved.
The central charge condition implies a restriction on the allowed maps $X^{i}$ to the compactified sector of the target space, such that the associated one-form decomposes into a harmonic one form $dX^i_h$ with integer coefficients (winding numbers) and an exact one $dA^i$ which represents the new dynamical degrees of freedom,
\begin{equation}
\small
\label{hodgedecomposition}
dX^i(\sigma^1, \sigma^2, \tau)= dX^i_h(\sigma^1, \sigma^2)+dA^i(\sigma^1, \sigma^2,\tau)
\end{equation}

The Hamiltonian formulation of the $D=11$ Supermembrane with irreducible winding (or nontrivial central charge) was found in  \cite{Ovalle3}  
\begin{equation}
\small
\begin{aligned}
\label{hamiltonianirred10}
H&=\int_\Sigma d^2\sigma\sqrt{w}\Big[\frac{1}{2}\Big(\frac{P_m}{\sqrt{w}}\Big)^2+\frac{1}{2}\Big(\frac{P_i}{\sqrt{w}}\Big)^2 + \frac{1}{4}\left\{X^m,X^m\right\}^2 + \frac{1}{2}(\mathcal{D}_iX^m)^2+\frac{1}{4}(\mathcal{F}_{ij})^2 \Big] \\
&+ \int_\Sigma d^2\sigma\sqrt{w}\Big[\Lambda\Big(\mathcal{D}_i\big(\frac{P_i}{\sqrt{w}}\big)+\left\{X^m,\frac{P_m}{\sqrt{w}}\right\} \Big)\Big] +(n^2Area_{T^2}^2)\\
&+ \int_\Sigma d^2\sigma\sqrt{w}\Big[-\bar{\theta}\Gamma_-\Gamma_i\mathcal{D}_i\theta-\bar{\theta}\Gamma_-\Gamma_m\left\{X^m,\theta\right\}+\Lambda\left\{\bar{\theta}\Gamma_-,\theta\right\}\Big]\,,
\end{aligned}
\end{equation}
where there is a symplectic covariant derivative and symplectic curvature defined
\begin{equation}
\label{symplecticfields}
\mathcal{D}_iX^m = D_iX^m+\left\{ A_i,X^m\right\}, \qquad  \mathcal{F}_{ij}= D_iA_j-D_jA_i+\left\{ A_i,A_j\right\},
\end{equation}
with $D_i$ a covariant derivative defined in terms of the moduli of the torus \cite{mpgm2},
\begin{equation*}
   D_i\bullet=2\pi R _i m_i^k \theta_{kj} \frac{\epsilon^{uv}}{\sqrt{w}}\partial_u\hat{X}^j\partial_v\bullet= \left\{ {X}_h^j,\bullet \right\}\delta_{ij} \,,  
\end{equation*}

Classicaly the Hamiltonian does not contain string-like spikes \cite{mpgm}. At a quantum level it has  the remarkable property of having a supersymmetric discrete spectrum with finite multiplicity, \cite{Boulton}. Since the Hamiltonian of both theories, given by (\ref{hamiltonianirred10}), differ at most in a constant, arising from the flux term associated to ${C_{+}}$ in the Hamiltonian (\ref{hamiltonian-fluxes}), the spectrum of the supermembrane with fluxes generated by ${C_{\pm}}$ has also discrete spectrum with finite multiplicity. 

The effect of the ${C_{\pm}}$ background also produces a discrete shift in some components of the momentum of the supermembrane, and in the Hamiltonian density. Comparing with the original configuration variables $(X^a,P_a)$ and considering the total momentum of the supermembrane, we have \cite{mpgm6}
\begin{equation}
\small
P^0_{-}=\int_{\Sigma}(\hat{P}_{-}+C_{-}) d\sigma^1 \wedge d\sigma^2 = \hat{P}^0_{-}+ k_-\,,
\end{equation}
\begin{equation}
\small
P^0_{+}=\int_{\Sigma}(\hat{P}_{+}+C_{+}) d\sigma^1 \wedge d\sigma^2 =H + k_+\,. 
\end{equation}
In the rest of the paper we will present new results characterizing other physical and geometrical aspects of the supermembrane formulated on a $M_9\times T^2$ background with constant fluxes $C_{\pm}$.
%

\section{Supercharges algebra of the M2-brane with constant $C_{\pm}$ fluxes}
In this section we obtain the algebra of the M2-brane charges in the presence of constant flux background.  The algebra of supercharges of the M2-brane with nontrivial flux is an algebra with central charges. The M2-brane algebra in noncompact  11D in the LCG was formerly worked out in \citep{dwhn}. In our construction we will closely follow its notation.
The supercharges for the supermembrane in $M_{9} \times T^2$ formulated in the LCG can be written as 
\begin{eqnarray}
   Q^+ & = & \int d^2\sigma (2P^m\Gamma_m + 2P^i\Gamma_i + \sqrt{w}\{X^m,X^n\}\Gamma_{mn} \nonumber \\ & + & 2\sqrt{w}\{X^m,X^i\}\Gamma_{mi}+\sqrt{w}\{X^i,X^j\}\Gamma_{ij})\theta, \\
   Q^- & = & 2\Gamma^-\theta_0.
\end{eqnarray}
Thus, using the definitions of the zero modes
\begin{eqnarray}
P_0^m  =  \int_\Sigma d^2\sigma P^m, \quad  P^i_{KK}\equiv P_0^i  =  \int_\Sigma d^2\sigma P^i %
\end{eqnarray}
\begin{equation}
\hat{X}_0^m =  \int_\Sigma d^2 \sigma \sqrt{w(\sigma)}\hat{X}^m, \quad \hat{X}_0^i =  \int_\Sigma d^2 \sigma \sqrt{w(\sigma)}\hat{X}^i,  \quad \hat{\theta_0} =  \int_\Sigma d^2 \sigma \sqrt{w(\sigma)} \hat{\theta},
\end{equation}
where we denote $P^i_{KK}$ the zero mode momentum contribution associated with the compact directions in order to emphasize that these also represent the KK modes and therefore are constants.   
We can see, that $Q^-$ is already a zero mode contribution and on the other hand the $Q^+$ can be written as
\begin{eqnarray}\label{eqn 3.5}
    Q^+ &=& \int d^2\sigma (2P'^m\Gamma_m + 2P'^i\Gamma_i + \sqrt{w}\{X'^m,X'^n\}\Gamma_{mn}  \nonumber \\ &+& 2\sqrt{w}\{X'^m,X'^i\}\Gamma_{mi}+\sqrt{w}\{X'^i,X'^j\}\Gamma_{ij})\theta' \nonumber \\ & + & 2P_0^m  \Gamma_m \theta_0 + 2P_{KK}^i  \Gamma_i \theta_0  + \left(\int d^2\sigma \sqrt{w}\{X'^i,X'^j\}\right)\Gamma_{ij}\theta_0,  \label{q+}
\end{eqnarray}
where the primes indicates that we are excluding the zero mode contributions. Due to the last term in this equation we can see that in general  there will not be a decoupling of the zero modes in the $Q^+$, unless this term is a constant.

If we impose the flux condition (\ref{fluxcondition2}) over the form $C_{\pm}$ it corresponds to the M2-brane in the LCG compactified on a $M_9\times T^2$ target space subject to the irreducible wrapping condition (\ref{cc}). That is, all configurations must satisfy that the bracket on the later term of (\ref{eqn 3.5}) is proportional to $n$.
It implies the existence of a central charge contribution $n$ in the supersymmetric algebra. We can replace this integral by its constant expression. In fact its Poisson bracket with any functional is zero, since its Poisson bracket with $P^{'i}$ is zero.
 In the following we will specify the flux units  $(k_+,k_-)=(k_+,n)$ being $n$ the integer associated to the central charge.
The equation (\ref{q+}) takes then the following form
\begin{eqnarray}
\small
Q^+ & = & \int d^2\sigma (2P'^m\Gamma_m + 2P'^i\Gamma_i \sqrt{w}\{X'^m,X'^n\}\Gamma_{mn}+2\sqrt{w}\{X'^m,X'^i\}\Gamma_{mi} \nonumber \\ &+& \sqrt{w}\{X'^i,X'^j\}\Gamma_{ij})\theta' +[2P_0^m \Gamma_m  + 2P_{KK}^i \Gamma_i + \epsilon^{ij}n\Gamma_{ij}]\theta_0,  \label{q+2}
\end{eqnarray}
in which the zero mode contributions can now be separated. Then we can decouple the zero mode contribution to the charges
\begin{eqnarray}
Q^- = Q^-_0 & = & 2\Gamma^-\theta_0, \\
Q^+_0 & = & 2P_0^m \Gamma_m \theta_0 +  2P_{KK}^i \Gamma_i \theta_0 + \epsilon^{ij}n\Gamma_{ij}\theta_0.
\end{eqnarray}
Now we can compute the algebra of the zero modes to obtain 
\begin{eqnarray}
((Q^-_0)^\alpha,(Q_0^-)_\beta)_{DB} & = & -2(\Gamma^-)^\alpha_\beta \\
((Q^+_0)^\alpha,(Q_0^-)_\beta)_{DB} & = &-(\Gamma_m\Gamma^+\Gamma^-)^\alpha_\beta P_0^m -(\Gamma_i\Gamma^+\Gamma^-)^\alpha_\beta P_{KK}^i \nonumber \\ &-& \frac{1}{2}(\Gamma_{ij}\Gamma^+\Gamma^-)^\alpha_\beta \epsilon^{ij}n \\
((Q^+_0)^\alpha,(Q_0^+)_\beta)_{DB} & = & [(P_0)^2 + (P_{KK})^2 ](\Gamma^+)^\alpha_\beta + 2(\Gamma^+\Gamma^i)^\alpha_\beta P^j_{KK}\epsilon_{ij}n \nonumber \\ & + & (\Gamma^+)^\alpha_\beta n^2
\end{eqnarray}
The complete algebra of the zero modes supercharges is
\begin{eqnarray}
(Q_0^\alpha, Q_{0\beta})_{D.B} & = &
[(P_0)^2+(P_{KK})^2+n^2] (\Gamma^+)^\alpha_\beta-2(\Gamma_m)^\alpha_\beta P_0^m  \nonumber  \\ & + &2(\Gamma^+ \Gamma^i)^\alpha_\beta P^j_{KK}\epsilon_{ij}n-(\Gamma_{ij})^\alpha_\beta \epsilon^{ij}n-2 (\Gamma_i)^\alpha_\beta P_{KK}^i \nonumber \\ & - & 2(\Gamma^-)^\alpha_\beta .
\end{eqnarray}
The algebra of the supercharges without the zero mode contributions is
\begin{eqnarray}
((Q'^+)^\alpha,(Q'^+)_\beta)_{D.B} &=& (\Gamma^+)^\alpha_\beta (\mathcal{M}^2-n^2)-2(\Gamma_m\Gamma^+)^\alpha_\beta \int \sqrt{w}\varphi'X'^m \nonumber \\ & - & 2(\Gamma_i\Gamma^+)^\alpha_\beta \int \sqrt{w}\varphi'X'^i + (\Gamma_i\Gamma^+)^\alpha_\beta \int d^2\sigma \partial_uS'^{ui} \,,
\end{eqnarray}
where $\mathcal{M}^2$ is the mass operator and $\varphi$ is the first class constraint associated to the symplectomorphisms, or equivalently area preserving diffeomorphisms APD, the residual symmetry in the LCG,
\begin{equation}
\varphi'=\{(\sqrt{w})^{-1}P',X'\}+\{\Bar{\theta}'\Gamma^-,\theta'\}=0
\end{equation}
and the integrand of the surface term is
\begin{equation}
    S'^{ui}=\epsilon^{uv}X'^i\left(\frac{2}{\sqrt{w}}P'_m\partial_v X'^m + \frac{2}{\sqrt{w}}P'_j\partial_v X'^j + \Bar{\theta}'\Gamma^-\partial_v\theta'\right)
\end{equation}

By analyzing in more detail we can observe that
\begin{eqnarray}
    \int_\Sigma d^2\sigma \partial_uS'^{ui} & = & \int d^2\sigma 2X'^i\varphi' \nonumber \\ & + &  2\int_\Sigma dX'^i \wedge \left(\frac{1}{\sqrt{w}}P'_md X'^m+\frac{1}{\sqrt{w}}P'_jdX'^j+\bar{\theta}'\Gamma^-d\theta'\right) , \nonumber \\ && 
\end{eqnarray}
which can be rewritten as 
\begin{equation}
    \int_\Sigma d^2\sigma \partial_uS'^{ui} =  \int d^2\sigma 2X'^i\varphi' + 2\oint_{C_u} dX'^i \phi_v \epsilon^{uv}  + 2(P_{KK})_j\epsilon^{ij}n\,,
\end{equation}
where we are denoting 
$\phi_v$ the global APD first class constraint,
\begin{equation}
    \phi_v = \oint_{C_v}\left(\frac{1}{\sqrt{w}}P_md X^m+\frac{1}{\sqrt{w}}P_jdX^j+\bar{\theta}\Gamma^-d\theta \right)=0 \,.
\end{equation}
The maps from the basis to the $T^2$ on the target in order to be well defined must satisfy condition
\begin{equation}
  \oint_{C_u} dX'^i= M^i_u \,.
\end{equation}
Finally the superalgebra of supersymmetric charges for the nonzero modes is
\begin{eqnarray}
(Q'^\alpha,Q'_\beta)_{D.B}  &=&(\Gamma^+)^\alpha_\beta (\mathcal{M}^2+2k_+) -2(\Gamma_m\Gamma^+)^\alpha_\beta \int \sqrt{w}\varphi'X'^m \nonumber \\  & + & 2(\Gamma_i\Gamma^+)(M^i_u \epsilon^{uv}\phi_v+ 2(P_{KK})_j\epsilon^{ij}n) \,,
\label{anzm}
\end{eqnarray}
where we have just made explicit the role of the central charge brackets.

 In this construction there exists a minimal embedding state associated to the fluxes, a 1/2 BPS state. Also there is a quantized nontrivial Kaluza Klein momentum  associated to 1/2 BPS states, that contributes to the multiplet.
The M2-brane with fluxes can have $1/4$ or $1/2$ of the supersymmetry preserved, depending if the Kaluza Klein modes are turned on, or not, respectively.  This is in agreement with the analysis of the $N=2$ superalgebra in (\cite{AbouZeid}).
\section{M2-brane with  fluxes: symplectomorphism transformations}
In this section we characterize the symmetries of the theory by a study on the area preserving diffeomorphisms APD which in two dimensions are equivalent to symplectomorphims. We consider the symplectomorphims connected and not connected to the identity. We find new results in which the constant harmonic map  $X_h^r(\sigma^1,\sigma^2)$ and the  single-valued map $A^r$ play a distinguish role. We will show that their transformations under symplectomorphisms make explicit hidden symmetries of the Hamiltonian $H$ that we will explore in the following section.

The Hamiltonian of the Supermembrane with central charges formulated in the LCG besides its invariance under symplectomorphisms is invariant under two discrete symmetries associated to the monodromy of the fiber denoted formerly by the authors \cite{mpgm7} as $S_{U}$ duality\footnote{the S-duality part of the U-duality group for M2-brane theory compactified on a 2-torus.} and the base respectively contained in $SL(2,Z)$, the group of isotopy classes of symplectomorphims. One of the discrete group symmetries $SL(2,Z)$ is associated to the change in the basis of harmonic one-forms  of the worldvolume torus of the M2-brane. Hence they will be relevant in the definition of symplectomorphims  connected to the identity $Symp_0(\Sigma)$ and not connected to the identity, ${Symp}_G(\Sigma)$, respectively. 

It was shown in \cite{dwhn,dwmn} that the $Symp_0(\Sigma)$ group is a symmetry of supermembrane theory when the target space is a $11D$ Minkowski spacetime. It can also be shown, we will explicitly do it in the following sections, that M2-brane theory with central charges is invariant under the different isotopy classes of area preserving diffeomorphism, in particular under $Symp_0(\Sigma)$. As each symplectomorphism over the torus is isotopic equivalent to a linear symplectomorphism, the changing from different isotopy classes is given by a matrix $S\in Sp(2,Z)\approx SL(2,Z) $ relating the linear diffeomorphisms \cite{mpgm2}. See \cite{Kahn} for a rigorous proof.  We are going to analyze the action of ${Symp}_0(T^2)$
and ${Symp}_G(T^2)$ separately.
%
\subsection{ $Symp_0(\Sigma)$  Transformations}

The supermembrane theory is invariant under symplectomorphisms connected to the identity \cite{dwmn},
the infinitesimal parameter $\xi$ defines a closed one-form $d(\xi_v d\sigma^v)=0$ which locally can be expressed
\begin{equation}
\xi_{v}=\partial_v\xi \,,
\end{equation}
with $\xi$ being either a function globally defined and whose associated $d\xi$ is an exact one-form or a function not globally defined whose associated $d\xi$ is a closed but not exact one-form, that is a harmonic one-form.

Any functional $O$ of the canonical variables transform locally under ${Symp}_0(T^2)$ as 
\begin{eqnarray}
\label{operator}
\delta O = \left\lbrace O, <d\xi\wedge \Big(\frac{P_a}{\sqrt{w}}dX^a+\theta\Gamma^-d\theta\Big)>\right\rbrace_{PB} \,.
\end{eqnarray}
%
%

In the above expression,
\begin{eqnarray}
 <d\xi\wedge \frac{P_a}{\sqrt{w}}dX^a> \equiv \int_\Sigma  \left(d\xi\wedge \frac{P_a}{\sqrt{w}}dX^a\right)
\end{eqnarray}
Therefore, under symplectomorphism connected to the identity
\begin{eqnarray}
\delta_{\xi} X^a =  \left\lbrace \xi, X^a\right\rbrace , \quad
\delta_{\xi} P_a =  \sqrt{w}\left\lbrace \xi, \frac{P_a}{\sqrt{w}}\right\rbrace , \quad
\delta_{\xi} \theta =  \left\lbrace \xi, \theta\right\rbrace
\end{eqnarray}
where $a$ runs over the compact and non-compact indices, $a=(i,m)$ with $i=1,2$ and $m=3,\dots,9$.

Using the Hodge decomposition previously introduced in  (\ref{hodgedecomposition}), the transformation of the maps under symplectomorphisms are
\begin{eqnarray}\label{eqn 4.5}
\delta_{\xi} X^i = \delta X_h^i + \delta A^i =  \left\lbrace \xi, X^i\right\rbrace \,,
\end{eqnarray}
where $dX^i_h$ are the harmonic one-forms and $dA^i$ the exact ones. We can consider all different possible transformations for the harmonic and exact classes. In this transformation one may not necessarily preserve the harmonicity property. That is, under diffeomorphisms the harmonic one-forms in one coordinate system transform to harmonic one-forms in the new coordinate system, however we are considering here transformation of the geometrical objects on the same coordinate system which will be symmetries of the theory but not necessarily corresponding to diffeomorphisms. 

In one case we will consider the cohomological class $d[X^i_h]$ and  we define an equivalence class $[X^i_h]$. Since we are interested to introduce a $U(1)$ gauge symmetry, we are not interested in the pure harmonic one-forms but on the $[X^i_h]$ class of maps which has an associated unique curvature $\widehat{F}$. In other case we will preserve the symplectic structure of the theory under the  transformation (\ref{eqn 4.5}).

In the following we will consider all different cases, with $\xi$ to be a general parameter not necessarily globally defined, although the associated differential will always be well defined. 
\begin{itemize}
\item First case: 
\begin{equation}
\label{eqn 4.6}
 \delta [X_h^i] = \left\lbrace \xi, [X_h^i] \right\rbrace, \quad \text{and} \quad \delta A^i = \left\lbrace \xi, A^i  \right\rbrace \,.
 \end{equation}

We will construct a $U(1)$ connection one-form associated to the harmonic one-form $d X^i_h$ with curvature $\widehat{F}$. The class $[X^i_h]$ is defined as follows: all the maps are obtained by a symplectic deformation of $X^i_h$ preserving ${\widehat{F}}$. The infinitesimal deformation is given by $\delta X_h^i = \left\lbrace \mu, X_h^i 
\right\rbrace$ where $\mu$ is an infinitesimal parameter. We will show that the (\ref{eqn 4.6}) preserves the equivalence class. The one form connection describes the realization over the worldvolume of the monopole connection associated to the $U(1)$ principal fiber bundle induced by the constant fluxes $\widehat{F}$.
Associated to $A^i$ we will also introduce a one-form connection on a trivial $U(1)$ principal bundle which carries the dynamical degrees of freedom.
%

%
%
\item Second case: 
\begin{equation} 
\label{secondcasesec4}
\delta X_h^i = 0 \quad \text{and} \quad \delta A^i = \left\lbrace \xi, X^i  \right\rbrace 
\end{equation}   

In this second case the complete transformation acts only on the exact part of the embedding map $A$ leaving invariant the harmonic sector. This transformation is associated to a symplectic connection $A$ that carries the dynamical degrees of freedom. This transformation law and the symplectic connection has been extensively discussed in \cite{Ovalle3}.
\item Third case: 
\begin{equation} 
\delta [X_h^i] = \left\lbrace \xi, [X^i] \right\rbrace \quad \text{and} \quad \delta A^i = 0.
\end{equation} 
In this third case the symplectic transformation is completely associated to the harmonic sector. We did not find any new connection, expressed in terms of the maps defining the supermembrane, associated to this transformation. We will not discuss it any longer.
\end{itemize}
%
%
\subsection{The full group of  symplectomorphism transformations}
We consider in this section the full group of symplectomorphisms including the ones which are not connected to the identity. As we mentioned before, in real dimension 2 the group of symplectomorphisms coincides with the group of diffeomorphisms fixing a given volume form, APD. In higher dimensions both groups have very different topological properties. For any dimension the full group of diffeomorphisms of a smooth manifold is homotopy equivalent to the volume preserving diffeomorphisms, VPD. Consequently, this also occurs with the full group of diffeomorphisms and the symplectomorphisms in two dimensions.

The Supermembrane formulation in this work is in terms of a compact torus $\Sigma$ on the worldvolume and a flat torus $T^2$ on the compactified sector of the target space. The maps from $\Sigma$ to $T^2$ are scalar fields under symplectomorphisms connected to the identity on $\Sigma$ with values on $T^2$.  We will now discuss the transformation law under symplectomorphisms not connected to the identity.\newline

$T^2$ is characterized by its moduli: the Teichmm\"uller parameter $\tau$, $Im \tau > 0$ and a real $R$ radius. The maps to the compactified sector of the target space are defined as

\begin{equation}
\label{ec1notas}
p \in \Sigma \, \to \, \int_{p_o}^{p} dX^i \, \in \, \mathbb{C} \,, \text{the complex plane}\,; \quad i=1,2\,.
\end{equation}
where
\begin{equation}
\label{ec2notas}
\oint_{\mathcal{C}_j} d \left(X^1 + iX^2 \right)= 2 \pi R \left(l_j + m_j \tau \right) \, \in  \mathcal{L} \,.
\end{equation}
$\mathcal{C}_j $ is a basis of homology on $\Sigma$, $j=1,2$, $l_j, m_j $ are integers (the winding numbers) and $\mathcal{L}$ is a lattice on the complex plane $\mathbb{C}$, $T^2 \equiv \mathbb{C}/ {\mathcal{L}}$.
Associated to a given basis of homology there is a basis of harmonic 1-forms $\omega^i$, $i=1,2$, normalized by
\begin{equation}
\oint_{\mathcal{C}_j}  \omega^i = \delta^i_j \,.
\end{equation}
The harmonic 1-forms are closed $d\omega^i=0$ and coclosed $d* \omega^i=0$. Locally any closed 1-form can be expressed as 
\begin{equation}
\omega^i = d \widehat{X}^i \,, \quad i=1,2 \,,
\end{equation}
then $d * d\hat{X}^i=0$, $i=1,2$. In terms of the local coordinates $z= \sigma^1 + i \sigma^2$ on $\Sigma$ we have

\begin{equation}
\sum_a \partial_a^2\widehat{X}^i =0 \,, \quad i=1,2;a=1,2 \,.
\end{equation}
The most general closed 1-forms on $\Sigma$ can be expressed, using the Hodge decomposition in harmonic and exact 1-forms $dA^i$, as
\begin{equation}
\label{ec4nuevasnotas}
d X^i = M^i_j d \hat{X}^j + dA^i \,,
\end{equation}
$A^i$, $i=1,2$, are then single valued functions on  $\Sigma$. From (\ref{ec2notas}) and (\ref{ec4nuevasnotas}) we obtain
\begin{equation}
\label{ec15notas}
d \left(X^1 + iX^2 \right)= 2 \pi R \left(l_j + m_j \tau \right) d \hat{X}^j + d\left(A^1 + iA^2\right)\,\, .
\end{equation}
%


 Under the full group of symplectomorphisms, the 1-forms $dX^m$ remain invariant. We will obtain the transformation law for $dX^i$, $i=1,2$. Under a symplectomorphism connected to the identity on $\Sigma$, the homology basis $\mathcal{C}_j$ and the harmonic basis $d\hat{X}^i$ remain invariant. Under symplectomorphisms not connected to the identity, the homology basis transforms by the action of $SL(2,Z)$ and the harmonic basis by a corresponding $SL(2)$ transformation
\begin{equation}
\label{ec5notas}
d \hat{X}^i \, \to S^i_k d \hat{X}^k \,, \quad \, \mathcal{C}_j  \, \to (S^{-1})^l_j \, \mathcal{C}_l \,, \quad \textrm{with} \quad S \in SL(2,Z) \,.
\end{equation}
We notice that the map 
\begin{equation}
\label{bijection}
\int_{P_o}^{P} d \hat{X}^i \, : \, \Sigma \rightarrow T^2= \mathbb{C}/ {\mathcal{L}}   
\end{equation}
is an immersion, since $d \hat{X}^1 \wedge d \hat{X}^2$ is nondegenerate. Moreover, it is a bijection since it is surjective and the tori $\Sigma$ and $T^2$ are compact. We can then pullback and pushforward the symplectic structures on $T^2$ and $\Sigma$ by the bijection (\ref{bijection}). We will consider, from now on, the symplectomorphisms on $\Sigma$ and $T^2$ always related by (\ref{bijection}). A symplectomorphism connected or nonconnected to the identity on $T^2$ induces a symplectomorphism connected or non connected to the identity on $\Sigma$ and viceversa.
The flat torus $T^2$ is defined by the parameters $R$ and $\tau$. Under symplectomorphisms connected to the identity on $T^2$, $R$ and $\tau$ remain invariant. The corresponding symplectomorphism on $\Sigma$ leaves invariant the homology basis as well as the normalized basis of harmonics. The 1-forms $d{X}^i$, $i=1,2$ remain invariant and so does $d{X}^m$ \cite{mpgm3} and
\begin{equation}
\sqrt{w} \equiv \frac{1}{2} \epsilon_{ij} \partial_u \hat{X}^i \partial_v \hat{X}^j \epsilon^{uv} \,. 
\end{equation}
 The map (\ref{ec1notas}), (\ref{ec2notas}) is then invariant and consequently the Hamiltonian also is invariant. So the map (\ref{ec1notas}) y (\ref{ec2notas}) can be elevated to a map from $\Sigma$ to the flat torus $T^2$ modulo symplectomorphisms connected to the identity.

On $\Sigma$ the realization of the symplectomorphisms connected to the identity is generated by the first class constraint on the supermembrane Hamiltonian. We may also consider the action of symplectomorphisms non-connected to the identity. In this case \citep{mpgm2, mpgm3}

\begin{equation}
\label{tautransformation} 
{\tau} \rightarrow \frac{{{a}{\tau}}+{{b}} }{{{c}}{\tau}+{{d}}} \, \qquad, \, \left( \begin{array}{cc} a & b \\ c & d \end{array}\right) \in SL(2,{Z})
\end{equation}
hence 
\begin{equation}
\label{ec12notas}
Im \tau \, \to \frac{Im \tau }{|c \tau +d|^2}
\, ,
\end{equation}
and
\begin{equation}
\label{ec13notas}
R \to \, R\, |c \tau +d| \, ,
\end{equation}
since the volume of the $T^2$ is proportional to $R^2 Im\,\tau $ and the transformation is volume preserving (or area preserving in 2 dimensions). The non-connected to the identity transformation on $T^2$ induces via (\ref{bijection}) a transformation on $d \hat{X}^i$, $i=1,2$
\begin{equation}
d \hat{X}^i \, \to S^i_j d \hat{X}^j \,, \, \quad S \in SL(2,Z) \,.
\end{equation}
We then consider the transformation on the winding matrix to be:
\begin{equation}
\label{ec14notas}
\begin{pmatrix} m_1 & m_2 \\
                        l_1 & l_2
 \end{pmatrix} \to {\begin{pmatrix} a & c\\
                        b & d
 \end{pmatrix}}^{-1}\begin{pmatrix} m_1 & m_2 \\
                        l_1 & l_2
 \end{pmatrix} S^{-1}\,.
\end{equation}
The resulting transformation on $d{X}^i$ is
\begin{equation}\label{eqn 4.24}
d X^1 + idX^2 \to d \left(X^1 + iX^2 \right) e^{i\varphi} \, .
\end{equation}
where
\begin{equation}
\label{ec16notas}
\frac{c\tau+d}{|c\tau+d|}= e^{-i\varphi}\, \,.
\end{equation}
It turns out that the Hamiltonian is invariant under the change (\ref{eqn 4.24}), consequently it is invariant under the full symplectomorphism group preserving the canonical symplectic 2-form on $T^2$. The transformations (\ref{tautransformation}), (\ref{ec13notas}), (\ref{ec14notas}), (\ref{ec16notas}) play a relevant role in the U-duality invariance of the theory with central charge \cite{mpgm7}.
This result is important since the Supermembrane is then well defined on the class of $T^2$ flat torus modulo the full group of symplectomorphisms. Consequently, one may define a symplectic torus bundle with base manifold $\Sigma$, fiber $T^2$ and structure group the symplectomorphisms \cite{mpgm3} on $T^2$. In addition the nontrivial central charge introduces a nontrivial $U(1)$ bundle in the geometric structure and we we have to prove consistency of the overall construction.

In the previous section we considered infinitesimal symplectomorphism connected to the identity. Let $\beta_0$ be a symplectomorphism not connected to the identity, $\beta_0$ belongs to an isotopy class of symplectomorphisms. Let $\beta_1$ be on the same class. Then there exist a smooth family of symplectomorphisms ${h_t}$ on the same class such that $h_1=\beta_1$ and $h_0=\beta_0$. $\beta_1$ can always be expressed as $\beta_1=\beta_0({\beta_0}^{-1}h_1)$, where ${\beta_0}^{-1}h_t$ is a family of symplectomorphisms connected to the identity, since ${\beta_0}^{-1}h_0=\mathbb{I}$, and ${\beta_0}^{-1}\beta_1$ is then a symplectomorphism connected to the identity. Consequently any symplectomorphism on the same class of $\beta_0$ can be written as a product $\beta_0 f$, where $f$ is connected to the identity. Also $\beta_0 f=g \beta_0$ where g is also connected to the identity. In the case of the symplectomorphisms on a torus, the group generated by the isotopy classes is  $\pi_0(G)=SL(2,Z)$, as we have already used. It is the same as the $\pi_0 (\textit{Diff($T^2$)})$\footnote{Strictly speaking $\pi_0(Symp(T^2))=\pi_0(Diff^+(T^2))$, but in the case of the 2-torus $Diff(T^2)\approx Diff^+(T^2)$} since diffeomorphisms are homotopic to the volume preserving diffeomorphisms, and these ones with the symplectomorphisms on a 2 dimensional surface. We conclude that the infinitesimal transformations in the previous section can be also defined on the isotopy classes not connected to the identity by taking:
\begin{eqnarray}
&\textit{First case:} & \hat{X}^{'i}=S^i_j\hat{X}^j + S^i_j \left\lbrace \xi, \hat{X}^j \right\rbrace \, , \quad A^{'i}=A^{i}e^{i\varphi}+\left\lbrace \xi, {A}^i \right\rbrace \, .\\
&\textit{Second case:} & \hat{X}^{'i}=S^{i}_{j}\hat{X}^j\, , \quad A^{'i}=A^{i}e^{i\varphi}+\left\lbrace \xi, {X}^i \right\rbrace \, ,
\end{eqnarray}
where $\left\lbrace S^{i}_{j} \right\rbrace \in SL(2,Z)$. We have already dismissed the third case.
Notice that the transformation of the first class does not preserve the decomposition of the closed one-forms into harmonic and exact one-forms since we are in that case interested in $U(1)$ gauge equivalent classes. 

The main step now is to construct an $U(1)$ connection 1-form on $\Sigma$ which arises from a connection on a principle $U(1)$ bundle over $\Sigma$ compatible with the symplectic torus bundle.
%
%
%

\section{ Symmetries of the M2-brane theory with fluxes}
In this section we will show the existence of connection one-forms associated to the symmetries described in section 4. In our construction the torus $\Sigma$, the base manifold, and the flat torus $T^2$ on the target space are diffeomorphic. The immersion defined by $\int_{P_o}^{P} dX^i$ from $\Sigma \to T^2$ is also surjective and injective. Hence is a bijective map and since the symplectic two form $d\hat{X}^{'i} \wedge d\hat{X}^j \epsilon_{ij}$ is nondegenerate, it is a diffeomorphism between $\Sigma$ and $T^2$. This is relevant since we can then pullback and pushforward vector bundle from $\Sigma \leftrightarrow T^2$. In particular, the existence of a nontrivial central charge on $\Sigma$ is related to the existence of fluxes on the compactified sector of the target space and viceversa. That is, the nontrivial $U(1)$ principal bundle on $\Sigma$ associated to the central charge can be pushforward to a nontrivial $U(1)$ principal on $T^2$ associated to the flux condition and viceversa. In this section we will introduce
\begin{itemize}
    \item {A $U(1)$ connection 1-form associated to the central charge on $\Sigma$ or equivalently to the flux condition on the target. It is associated to the non-trivial $U(1)$ principle bundle with base manifold $\Sigma$, and characterized by the Chern number $n$. It is a monopole $U(1)$ connection, we denote it $\widehat{A}$.}
    \item{A $U(1)$ connection 1-form on the same $U(1)$ principle bundle which in addition to the topological structure associated to $\widehat{A}$, it carries the physical degrees of freedom associated to the compactified sector of the Supermembrane. We will denote it $\mathbb{A}$}. Its curvature satisfies
    \begin{equation}
    \label{ec1notasseccion5}
      \frac{1}{2 \pi} \int_{\Sigma} \mathbb{F} = n \,, 
    \end{equation}
hence it has the same  Chern number as $\widehat{F}=d \widehat{A}$. Consequently, it is a connection 1-form associated to the same nontrivial $U(1)$ principle bundle as $\widehat{A}$.
\item{A symplectic 1-form connection $A=A_i d\sigma^i$ on $\Sigma$ associated to a symplectic principle bundle with base $\Sigma$ and structure group $G$, the symplectomorphisms preserving the nondegenerate two form $\sqrt{w} \epsilon_{ij} d\sigma^i \wedge d\sigma^j$. This one is the pullback under the minimal map of the canonical symplectic 1-form on the flat $T^2$ torus on the target space. The curvature 2-form  on $\Sigma$, $\mathcal{F}$, satisfies}
    \begin{equation}
    \label{ec2notasseccion5}
      \frac{1}{2 \pi} \int_{\Sigma} {\mathcal{F}} = 0 \,. 
    \end{equation}
The connection 1-form $A$ carries the degrees of freedom of the compactified sector of the Supermembrane, but does not provide a monopole topological structure as in the previous cases.
\end{itemize}

We refer as first case or second case the ones already mentioned on the previous section and defined by two different gauge symmetries:\newline
\begin{itemize}
\item{{\bf U(1) Gauge Symmetry}: Let us consider the infinitesimal symplectomorphism transformation previously discussed in (\ref{eqn 4.6})

\begin{equation}
\label{ec5notasseccion5}
\delta [X_h^i] = \left\lbrace \xi, [X_h^i] \right\rbrace,\quad\delta A^i = \left\lbrace \xi, A^i \right\rbrace \quad \text{and} \quad i=1,2 \,.
\end{equation}

In the following we will show the appearance of a gauge connection $\mathbb{A}$ composed by two different one-form connections $\widehat{A}$ and $\mathcal{A}$ that transform differently under the previous symplectomorphism transformation
 We define $\widehat{A}=\frac{1}{2}\epsilon_{ij}X_h^idX_h^j$. It is not a global 1-form on $\Sigma$ but a connection 1-form on $\Sigma$. In fact, notice that $X_h^i$ is not a singled valued function on $\Sigma$. However, under  (\ref{ec5notasseccion5}),
$\widehat{A}$ transforms as a $U(1)$ gauge vector \cite{Restuccia}
\begin{eqnarray}
\label{transfAtechoeta}
 \delta \widehat{A} =d \eta \,, \quad \eta = -\frac{\epsilon^{uv}}{\sqrt{w}}\partial_v\xi \widehat{A}_u -\xi\star \widehat{F} \,,
\end{eqnarray}
where $\star \widehat{F}$ is the Hodge dual of the two form $\widehat{F}=d \widehat{A}$. It satisfies, by definition of $\sqrt{w}$, $\star \widehat{F}=n$. The curvature $\widehat{F}= \frac{1}{2}\epsilon_{ij}dX_h^i \wedge  d X_h^j$  is a closed 2-form satisfying
\begin{equation}
    \label{integralflujoT2}
      \frac{1}{2 \pi} \int_{T^2} \widehat{F} = n \,, 
    \end{equation}
where we have normalized the area of $T^2$ to 1. Consequently $\widehat{F}$ is the curvature of a connection 1-form of a nontrivial $U(1)$ principle bundle characterized by the Chern number $n$. The same results (\ref{transfAtechoeta}) and (\ref{integralflujoT2}) occur for the variation and curvature associated to each member of the class, where we replace in $\widehat{A}$, $X^i_h$ by the member of the class.

The gauge transformation generated by the infinitesimal transformation (\ref{eqn 4.6}) is associated not to the harmonic fields but to the equivalence class constructed from them. One of its elements is $X_h^i$ , $i=1,2$, but the other members are not harmonics, although they give rise to the same curvature $\widehat{F}$, which characterizes the equivalence class.\newline
We introduce a new one-form on $\Sigma$, not considered previously that also transforms under (\ref{eqn 4.6}),
\begin{equation}
\label{ec6notasseccion5}
\mathcal{A}=\frac{1}{2}\epsilon_{ij}(A^idX_h^j-A^jdX_h^i+A^idA^j) \,.    
\end{equation}
We notice that $\mathcal{A}$ is indeed a  1-form on $\Sigma$, it has an associated 2-form $\mathcal{F}^{U(1)}= d \mathcal{A}$ satisfying
    \begin{equation}
      \frac{1}{2 \pi} \int_{\Sigma} \mathcal{F}^{U(1)} = 0 \,, 
    \end{equation}

Under the infinitesimal transformation, (\ref{ec5notasseccion5}) we obtain 
\begin{equation}
\delta \mathcal{A} = d \widetilde \eta \, ,
\end{equation} 
\begin{equation}
\widetilde \eta\equiv \left( -\frac{\epsilon^{uv}}{\sqrt\omega} \partial_{v}\xi (\frac{1}{2}\epsilon_{{i}{j}}A^i\partial_{u}X_h^j) -\xi * \widehat F \right) \,.
\end{equation} 
It is important to remark that although $\mathcal{A}$ behaves as an exact $U(1)$ connection, physically it is relevant since it carries the information associated to the dynamical degrees of freedom of the theory $A^i(\sigma^1,\sigma^2,\tau)$.

Now it is possible to define the following one-form  linear combination  $\mathbb{A} \equiv \widehat{A}+ \lambda \mathcal{A}$. It transforms under the complete infinitesimal transformation (\ref{eqn 4.6}) as
\begin{equation}
\delta \mathbb{A} = d(\eta + \lambda \widetilde \eta) \,\,, \quad \lambda \,\,  \text{a real constant} \,,
\end{equation}
a $U(1)$ connection 1-form with curvature $\mathbb{F}=d\mathbb{A}$ satisfying
    \begin{equation}
      \frac{1}{2 \pi} \int_{\Sigma} \mathbb{F} = n \,, 
    \end{equation}
hence $\mathbb{A}$ is a connection one form on the same $U(1)$ principle bundle with Chern number $n$. $\mathbb{A}$ besides the topological structure provided by $\widehat{A}$, it carries the physical degrees of freedom associated to $A^i$ the single valued fields describing the maps}. 
\item{{\bf Symplectic Gauge Symmetry}: We consider the second class of infinitesimal transformation considered in (\ref{secondcasesec4}), $\delta X_h^i = 0$ and $\delta A^i = \mathcal{D}_i \xi$, where $\mathcal{D}_i \equiv \left\lbrace \bullet, X_h^i  \right\rbrace + \left\lbrace \bullet, A^i  \right\rbrace$ is a covariant derivative which satisfies the Leibnitz rule and preserves the transformation law of its argument under symplectomorphisms. That is, if $\delta X^m = \left\lbrace \xi, X^m  \right\rbrace$ then 
\begin{equation}
\delta \mathcal{D}_i X^m = \left\lbrace \xi, \mathcal{D}_i X^m \right\rbrace \,.
\end{equation}
The above transformation law corresponds to an infinitesimal symplectomorphism, connected to the identity composed with a transformation within the cohomology class of ${X}^i_h$, under which the harmonic basis is invariant. $A^i$ has the transformation law of a symplectic connection 1-form on $\Sigma$ with curvature $\mathcal{F}_{ij}=D_iA_j-D_jA_i+\{A_i,A_j\}$ \cite{Ovalle3}. We notice that $\mathcal{F}_{ij}$ is a total derivative, hence 
    \begin{equation}
      \frac{1}{2 \pi} \int_{\Sigma} \mathcal{F} = 0 \,. 
    \end{equation}
Although, $A=A_u d\sigma^u$ carries the physical degrees of freedom of the maps to the compact sector on target space, the monopole structure is missing on this bundle.
}

A not trivial property of the $U(1)$ curvature $\mathcal{F}^{U(1)}=d\mathcal{A}$ and the symplectic curvature $\mathcal{F}$ is they are the same when expressed in terms of its components fields $\widehat{X},A^i,$ i.e.
$$\mathcal{F}^{U(1)}=\mathcal{F}=D_iA_j-D_jA_i+\{A_i,A_j\}$$
This result will be relevant in subsection 5.2.
\item{\bf Global $U(1)$ symmetry}

A global symmetry is induced in the Hamiltonian by the infinitesimal symplectomorphims action connected to the identity on the target torus $T^2$:
\begin{equation}
\widetilde{X}^i\to \widetilde{X}^i +\xi^i(T^2)
\end{equation} \,,
preserving the torus area $\partial_i(\sqrt{W_{T^2}}\xi^i(\widetilde{X}))=0$ where $\widetilde{X}$ represents the $T^2$ coordinates
where 
\begin{equation}
\xi^i=\frac{\epsilon^{ij}}{\sqrt{W_{T^2}}} \partial_j\alpha(\widetilde{X})
\end{equation}
and $\sqrt{W_{T^2}}=1$.

By using the definition $\widehat{A}=\frac{1}{2}\epsilon_{ij}\widehat{X}^id{\widehat{X}}^j$ identifying the harmonic maps $\widehat{X}^i$ with the coordinates of the target 2-torus $\widetilde{X}^i$, imposing that it should act as a one-form connection on $\Sigma$,
\begin{equation}
\label{transformationAtecho}
\delta\widehat{A}=\frac{1}{2}d(X^i\partial_i\alpha)) \,,
\end{equation}

we obtain that the parameter  $\alpha=\sum_i\lambda_i\widetilde{X}^i$ with $\lambda_i\in \mathbb{R}$. It is then a global $U(1)$ gauge transformation associated to a constant shift on the harmonic sector,
$$\widehat{X}^i\to \widehat{X}^i+\epsilon^{ik}\lambda_k.$$
This is a transformation of the harmonic sector into itself which leaves invariant the Hamiltonian. 
\end{itemize}
\subsection {Hamiltonian Symmetries}
The Hamiltonian of the Supermembrane is invariant under the full group of symplectomorphisms. The ones connected to the identity and the non-connected to the identity. The connected ones are generated by the first class constraint, the residual symmetry generator in the LCG. It generates the transformation on the Hamiltonian $H$ 
\begin{equation}
\delta {H}= \left\lbrace <\xi \phi>, H\right\rbrace_{P.B}\sim 0 \,,
\end{equation}
which is weakly zero. Besides the Hamiltonian is invariant under the non-connected to the identity symplectomorphisms. In fact, as discuss in section 4.2, the basis of homology transforms under $SL(2,Z)_{\Sigma}$ and so does the normalized basis of harmonic 1-forms. This transformation is pushed-forward to a transformation on the Teichm\"uller parameter $\tau$ on the target torus $T^2$, together with a transformation of the radius $R$ and the winding matrix, the $S_U$ duality. This last transformation generates a global $U(1)$ on the fields $X^i$, $i=1,2$ (\ref{ec15notas}) which leaves invariant the Hamiltonian. Moreover, the Hamiltonian is also invariant under a second $U(1)$ global symmetry (\ref{transformationAtecho}) generated by the $Symp_0(T^2)$. The Hamiltonian is invariant under all these transformations. \newline

The symplectomorphisms induce a local $U(1)$ transformation which becomes manifest by the presence of a $U(1)$ connection one form on $\Sigma$, denoted $\mathbb{A}$, with curvature $\mathbb{F}$. This connection is associated to a nontrivial $U(1)$ principle bundle which becomes physically relevant because it carries on one side the monopole structure associated to the nontrivial central charge and on the other side the dynamical fields associated to the compact sector of the M2-brane. The symplectomorphims also induce a symplectic  connection with symplectic curvature. The Hamiltonian is invariant under both gauge symmetries, the $U(1)$ and the symplectic one. In fact, the invariant Hamiltonian (\ref{hamiltonianirred10}) can be written as
\begin{equation}
\small
\begin{aligned}
\label{hamiltonianirred1}
H &=\int_\Sigma d^2\sigma\sqrt{w}\Big[\frac{1}{2}\Big(\frac{P_m}{\sqrt{w}}\Big)^2+\frac{1}{2}\Big(\frac{P_i}{\sqrt{w}}\Big)^2 + \frac{1}{4}\left\{X^m,X^m\right\}^2 + \frac{1}{2}(\mathcal{D}_iX^m)^2+\frac{1}{4}(\mathcal{F}_{ij})^2 \Big]\\&+\frac{1}{4}\int_\Sigma d^2\sigma\sqrt{w} \widehat{F}_{ij}^2 
+\int_\Sigma d^2\sigma\sqrt{w} \int_\Sigma d^2\sigma\sqrt{w}\Big[\Lambda\Big(\mathcal{D}_i\big(\frac{P_i}{\sqrt{w}}\big)+\left\{X^m,\frac{P_m}{\sqrt{w}}\right\} \Big)\Big] +\\
&+ \int_\Sigma d^2\sigma\sqrt{w}\Big[-\bar{\theta}\Gamma_-\Gamma_i\mathcal{D}_i\theta-\bar{\theta}\Gamma_-\Gamma_m\left\{X^m,\theta\right\}+\Lambda\left\{\bar{\theta}\Gamma_-,\theta\right\}\Big]\,,
\end{aligned}
\end{equation}
where the symplectic curvature $\mathcal{F}$ defined in (\ref{symplecticfields}), $\mathcal{F}_{ij}= D_iA_j-D_jA_i+\left\{ A_i,A_j\right\}$, appears explicitly. The Hamiltonian also admits an expression in which the $U(1)$ geometrical structure becomes manifest, showing the coupling to a nontrivial Maxwell density that contains also the flux contribution. 
This expression is obtained by replacing the curvature terms $\frac{1}{4}(\mathcal{F}_{ij})^2+\frac{1}{4}(\widehat{F}_{ij})^2$ in (\ref{hamiltonianirred1}) by $\frac{1}{4}({{\mathbb{F}}_{uv}})^2$,the Maxwell density Lagrangian of the $U(1)$ connection $\mathbb{A}$, that is,
\begin{equation}
\small
\begin{aligned}
\label{hamiltonianirred2}
H &=\int_\Sigma d^2\sigma\sqrt{w}\Big[\frac{1}{2}\Big(\frac{P_m}{\sqrt{w}}\Big)^2+\frac{1}{2}\Big(\frac{P_i}{\sqrt{w}}\Big)^2 + \frac{1}{4}\left\{X^m,X^m\right\}^2 + \frac{1}{2}\{X^i,X^m\}^2+\frac{1}{4}({{\mathbb{F}}_{uv}}{\mathbb{F}}^{uv}) +... \,,
\end{aligned}
\end{equation}
Both expressions in the Hamiltonian become equal. In fact,
\begin{equation}
{\mathcal{F}}^{ij}= \left\lbrace X_h^i, A^j \right\rbrace - \left\lbrace X_h^j, A^i \right\rbrace + \left\lbrace A^i, A^j \right\rbrace = \frac{1}{2} {\epsilon}^{ij} \frac{ {\epsilon}^{uv}}{\sqrt{w}} \mathcal{F}_{uv}=\frac{1}{2}{\epsilon}^{ij} \star{\mathcal{F}}\,,
\end{equation}
where 
\begin{equation}
{\mathcal{F}}_{uv}= {\epsilon}_{ij} (\partial_u X_h^i \partial_v A^j - \partial_u X_h^j \partial_v A^i+\partial_u A^i \partial_v A^j)\,.
\end{equation}
Also
\begin{equation}
{\widehat{F}}^{ij}= \left\lbrace X_h^i, X_h^j \right\rbrace  = \frac{1}{2} {\epsilon}^{ij} \frac{ {\epsilon}^{uv}}{\sqrt{w}} \widehat{F}_{uv}=\frac{1}{2}{\epsilon}^{ij} \star{\widehat{F}}\,,
\end{equation}
where
\begin{equation}
{\widehat{F}}_{uv}= {\epsilon}_{ij} \partial_u X_h^i \partial_v X_h^j \,.
\end{equation}
Then $\mathbb{F}= d \mathbb{A} = d\widehat{A} + d\mathcal{A}$, see section 5, satisfies
\begin{equation}
{\mathbb{F}}_{uv}= {\mathcal{F}}_{uv} + {\widehat{F}}_{uv} \quad , \star{\mathbb{F}}= \star{\mathcal{F}} + \star{\widehat{F}} \,,
\end{equation}
and, using that $\star{\widehat{F}}$ is constant independent of $\sigma$, together with (\ref{ec2notasseccion5})
\begin{equation}
\int_\Sigma d\sigma ^1 \wedge d\sigma ^2 \sqrt{w}\frac{1}{4}\Big[(\mathcal{F}^{ij})^2 + (\widehat{F}^{ij})^2 \Big] =\frac{1}{4} \int_\Sigma \mathbb{F} \star \mathbb{F} \,,
\end{equation}
where $\mathbb{F}=\frac{1}{2} \mathbb{F}_{uv}d\sigma ^u \wedge d\sigma ^v$. That is, the Maxwell action.
In this expression it becomes manifest the $U(1)$ dynamical curvature containing nontrivial topological information. We notice that $\mathcal{F}=d\mathcal{A}$ is either a $U(1)$ curvature on a trivial $U(1)$ principle bundle or $\mathcal{F}_{ij}= D_iA_j-D_jA_i+\left\{ A_i,A_j\right\}$ a symplectic curvature of the symplectic connection $A_i$. In this way both geometrical structures become directly related.
%
%
%

\section{Geometrical interpretation: A M2-brane on a twisted torus bundle} 
The central charge condition is associated to  the existence of a nontrivial $U(1)$ bundle associated to the presence of monopole configurations over the worldvolume. On the other hand the supermembrane compactified on $M_9\times T^2$ over a toroidal worldvolume can be extended to a formulation on a torus bundle over a torus with the target space geometry being the fiber. 
It is known that this global description is given in terms of symplectic torus bundles with monodromy in $SL(2,Z)$ classified according to the inequivalent coinvariant classes for a given monodromy.  Now we want to determine if there exists a relation between the $U(1)$ and symplectic bundle both of them in terms of the M2-brane fields over the worldvolume. We will show that this relation exists and moreover it becomes manifest when the M2-brne is formulated on symplectic twisted torus bundle. 
Let us consider the M2-brane global description in terms of the symplectic 2-torus bundle
\begin{eqnarray}
T^2 \rightarrow E \rightarrow \Sigma \, ,\hspace{0.5cm} G=Symp(T^2)
\end{eqnarray}
where $G$ is the structure group of the fiber. It was shown in \cite{mpgm3} that the M2-brane bundles have a monodromy defined as
\begin{eqnarray}
\mathcal{M} : \Pi_1(\Sigma) \rightarrow \Pi_0 (Symp(T^2)) = SL(2,Z) \,. 
\end{eqnarray}
Associated to the presence of the central charge condition there exists a $U(1)$ principal bundle fibered over $\Sigma$.
\begin{eqnarray}
U(1)\rightarrow E' \rightarrow \Sigma
\end{eqnarray}
A particularity of this $U(1)$ fiber is that its connection $\mathbb{A}$ is constructed in terms of the embedding maps of the M2-brane over the 2-torus target space, $X^i(\sigma^1,\sigma^2,\tau): \Sigma\to T^2$, with $i=1,2$ and
$\mathbb{A}= \widehat{A} + \mathcal{A}$, see section 5.
where $dX^i=dX_h^i+dA^i$  due to the Hodge decomposition. Relevantly the $dX_h^i$ do not have dependence on time, a crucial aspect towards its quantization. It gives rise to a nontrivial connection over the base manifold generalizing the notion of Dirac monopoles to Riemann surfaces of genus equal or larger than one as discussed in \cite{Restuccia2}. 

\subsection{Twisted 3-Torus}
In order to understand if there is a relation between the symplectic structure of the M2-brane and the principal $U(1)$ bundle, let us notice the existence of different twisted torus structures within M2-brane global description. Let us consider a twisted torus on the target. For example, the 2-torus of the target with local coordinates $\widetilde{X}^i, i=1,2$ with a nontrivial flux as described in section 2. 
If we denote $y$ the coordinate over the $S^1$ related to the principal bundle, we can define
\begin{eqnarray}
e^1&=&d\widetilde{X}^1 , \\
e^2&=&d\widetilde{X}^2 , \\
e^3&=&dy + n\widetilde{X}^1d\widetilde{X}^2
\end{eqnarray}
such that it satisfies the Maurer Cartan equation.
\begin{eqnarray}
de^3=f^3_{12}e^1\wedge e^2
\end{eqnarray}
for $f^3_{12}=n$, which is the structure equation related to a twisted torus $\mathbb{T}_W^3$. This geometrical structure can be pullback to the worldvolume torus $\Sigma$. We thus also have a twisted torus on $\Sigma$. Moreover, the twisted torus can be globally understood as a principal $U(1)$ bundle over the $T^2$ (the T-dual of the target torus with a nontrivial flux) or either a $T^2$ fibered over $S^1$ with parabolic monodromy in $SL(2,Z)$, identified by the integer characterizing the quantized flux.  
Therefore, there is another twisted  torus structures within the M2-brane bundle description. That is, a twisted torus  constructed from a  symplectic torus bundle over a homological one-cycle defined on the base $\Sigma$, the Maurer-Cartan equations are also satisfied. 

 In the twisted torus we consider, the presence of a $U(1)$ connection one-form constructed from the fields defining the M2-brane will be relevant. Hence we have a twisted torus with a connection one-form which characterizes the $U(1)$ principal bundle.\newline 
 It is important to mention that in this case the structure group is $Symp(T^2)\subset Diff^+(T^2)$ and not $Diff^+(T^2)$ as in the standard twisted torus. It is important to notice that in both cases the monodromy is contained in $SL(2,Z)$ because the isotopy classes of both are isomorphic to $SL(2,Z)$,
\begin{eqnarray}
\mbox{MPG}&=&\Pi_0(Diff^+(T^2))\approx SL(2,Z) , \\
\mbox{SMPG}&=&\Pi_0(Symp(T^2))\approx SL(2,Z).
\end{eqnarray}

\subsection{The Supermembrane on a twisted torus bundle} 
In \cite{mpgm3} we introduce the formulation of the Supermembrane on a symplectic torus bundle:
\begin{equation}
\label{ec1notasseccion6.2}
{T}^2 \rightarrow E \rightarrow \Sigma \, ,  
\end{equation}
with structure group the symplectomorphisms preserving the canonical symplectic 2-form on $T^2$. This symplectic structure can be pullback to the symplectomorphisms on $\Sigma$ preserving the symplectic 2-form induced by the diffeomorphism, defined in section 4, 
\begin{equation}
\label{ec2notasseccion6.2}
\Sigma \, \leftrightarrow \, \mathbb{C}/ {\mathcal{L}} \equiv T^2 \,,
\end{equation}
\begin{equation}
\label{ec3notasseccion6.2}
p \in \Sigma \, \rightarrow \, \small{(\int_{p_o}^{p} d\widehat{X}^i)} \, / {\mathcal{L}} \in T^2 \,.
\end{equation}
 There is a natural way to introduce the monodromy on the torus bundle as mentioned in 6.2. It is associated to the representations of the fundamental group of $\Sigma$ into the group of isotopy classes of symplectomorphisms on $T^2$, the group $SL(2,Z)$, which acts naturally on the first homology group of $T^2$.  It is then relevant for consistency the invariance of the Hamiltonian under the complete group of symplectomorphisms. This is so, because under symplectomorphisms connected to the identity the symplectic connection transforms as $\delta A_i= D_i\xi$ and the curvature as $\delta \mathcal{F}=\{\mathcal{F},\xi\}$. Hence the corresponding variations in the Hamiltonian are total derivatives. Also  as discussed in section 4.2 and section 5.1 under symplectomorphisms not connected to the identity the complex maps  $X_h^1+iX_h^2$ and $A_1+iA_2$ transform by a multiplicative phase. It turns out that the symplectic connection as well as the terms involving $X_h^i$ in the Hamiltonian are invariant under this transformation. In section 5, we introduced a class of maps, constructed from the harmonic one-forms, related by infinitesimal transformations connected to the identity
\begin{equation}
\label{ec4notasseccion6.2}
[\hat{X}^i] \, \rightarrow \, [\hat{X}^i +\{\xi,d\hat{X}^i\} ] \,,
\end{equation}
each element of the class have associated a $U(1)$ connection one-form $\widehat{A}$ which   transforms under (\ref{ec4notasseccion6.2}) as
\begin{equation}
\label{eqn 6.14}
\widehat{A} \, \rightarrow \,  \widehat{A} + d\eta \,.
\end{equation}
Hence $\widehat{A}$ remains in the same gauge equivalence class. We also introduced a 1-form $\mathcal{A}$ such that under
\begin{equation}
\label{ec5notasseccion6.2}
A \, \rightarrow \, A +\{\xi,A \}  \,,
\end{equation}
it transforms as
\begin{equation}
\label{eqn 6.16}
\mathcal{A} \, \rightarrow \, \mathcal{A} + d \widetilde{\eta}  \,.
\end{equation}
Moreover, $\mathbb{A}= \widehat{A} + \mathcal{A}$ is a connection one-form on a non-trivial $U(1)$ principal bundle and its Chern number, associated to the integral on $\Sigma$  of its curvature, corresponds to the quantized flux on the target, as shown in section 5.2. The connection carries not only the information of the non-trivial transitions on the $U(1)$ bundle, but also the information of the dynamical fields associated to the compact sector of the supermembrane. In addition to a twisted torus we have a connection on it. We can then define a geometry on the twisted torus preserving the bundle structure and we can couple backgrounds fields to the dynamics of the compact sector of the supermembrane.

The symplectomosphisms connected to the identity induces a $U(1)$ transformation, on the $U(1)$ connection, given by (\ref{eqn 6.14}) and (\ref{eqn 6.16}), leaving invariant its curvature. Also under symplectomorphisms not-connected to the identity on the $U(1)$ connection remains invariant.  Moreover the Hamiltonian can be expressed in terms of a symplectic connection and curvature or equivalently in terms of a $U(1)$ connection and its curvature making manifest the invariance under both group of transformations. 
We then have two gauge structures, the symplectic one associated to the structure group of the bundle and a $U(1)$ principal bundle associated to the central charge on the base $\Sigma$ or the flux condition on the target, both realized in terms of the physical degrees of freedom of the M2-brane.
The main point is that both are compatible. Under the monodromy
\begin{eqnarray}
\mathcal{M} : \Pi_{1} (\Sigma) \rightarrow \Pi_0 (Symp(T^2)) = SL(2,Z) \,,
\end{eqnarray}
on the symplectic torus bundle the symplectic isotopy classes transforms under $SL(2,Z)$ which induce a transformation of the fields describing the supermembrane, given in section 4, leaving the Hamiltonian invariant. Besides, the Hamiltonian can be re-expressed as in section 5.2 showing also invariance under the induced $U(1)$ transformation. So both geometrical structures are compatible, a non trivial property of the M2-brane theory. This means that the compact sector of the M2-brane can be realized in terms of the geometrical objects naturally defined on a twisted torus bundle
\begin{equation}
\label{ec7notasseccion6.2}
\mathbb{T}_W^3\equiv {T}_{U(1)}^2 \rightarrow E' \rightarrow \Sigma \, ,  
\end{equation}
 ${T}_{U(1)}^2$ denotes the twisted torus described by a $U(1)$ principle bundle over the flat torus on the target space. That is, the flat torus with a nontrivial flux. This geometric structure is represented by a connection one-form whose pullback to the worldvolume is the connection $\mathbb{A}$ we have introduced. The transition on the torus bundle represented by the monodromy is compatible with the transformation law of the connection, as we have shown. 

The relevance of this new geometrical interpretation in terms of a M2-brane twisted torus bundle is that it gives a definitive answer to the inequivalent classes of M2-brane bundles that exist for a M2-brane with central charges when it is compactified on a $M_9\times T^2$ target space. They can be classified by the monodromies of a twisted torus bundle on a torus, which are given by the coinvariants of the monodromy subgroups labelled with the charge of the quantized flux.

\section{Discussion and Conclusions}
The M2-brane compactified on $M_9\times T^2$ with $C_{\pm}$ fluxes is equivalent (modulo a constant shift) to a  supermembrane on the same target space subject to a   central charge condition associated to an irreducible wrapping condition and consequently the theory exhibits discrete spectrum. The so-called 'central charge'  condition is equivalent to have a nontrivial $U(1)$ principal bundle over the M2-brane worldvolume.

The algebra of supercharges is obtained , and it is shown that the zero modes decouple from the nonzero ones. We also find the amount of supersymmetry preserved by the theory. Since the constant fluxes imply the existence of a nontrivial central charge and there may be non vanishing Kaluza Klein states, -both of them BPS states breaking $1/2$ of supersymmetry-, then there are two possible multiplets: If the Kaluza Klein momentum state is turn on, the theory preserves $1/4$ of the original supersymmetry. If not the theory preserves $1/2$ of it.

Focusing on the M2-brane bundle description where the 2-torus target space is the fiber, with structure group the symplectomorphisms preserving its canonical symplectic two-form, and the worldvolume is the base manifold, where it is a formulation of the M2-brane on a torus bundle with monodromy in $SL(2,Z)$ as realized in \cite{mpgm7}. In this paper we show that the nontrivial $U(1)$ principal bundle over the base manifold and the 2-torus fiber determine a 3-twisted torus bundle $\mathbb{T}_W^3$ or equivalently a $T^2$ with a connection $U(1)$  that is consistently fibered over the worldvolume base. The $U(1)$ fiber as it is associated to a nontrivial flux condition it should not be interpreted as an extra spacetime dimension. We introduce a  connection one-form over the worldvolume. It is a  dynamical and topologically nontrivial $U(1)$ gauge field $\mathbb{A}$, compatible with the  symplectomorphisms transformations and with the transition on the torus bundle, given in terms of the monodromy.

The M2-brane with central charge realizes as symmmetries of the theory not only symplectomorphims connected with the identity -as it happens when there is no central charge but also those not connected with the identity. We provide the form of such general transformation including both sectors. This symmetry implies that the theory contains an extra $SL(2,Z)$ symmetry as formerly identified in \cite{mpgm3}, that plays a relevant role with respect its $U$-dual invariance as discussed in \cite{mpgm7}. 
We find new $U(1)$ symmetries (gauge and global) of the M2-brane worldvolume theory:  There is a new dynamically nontrivial $U(1)$ symmetry with gauge connection $\mathbb{A}$ that contains a one-form connection $\widehat{A}$  associated to the constant flux 2-form $\widehat{F}$ curvature, and  a dynamical single-valued one $\mathcal{A}$  topologically trivial whose curvature is $\mathcal{F}$. There exists also a symplectic symmetry with connection $A$ whose associated symplectic curvature is also $\mathcal{F}$.  Because of this nontrivial property, the  Hamiltonian of the Supermembrane with constant fluxes $C_{\pm}$ can exhibit both types of symmetries: it describes a M2-brane with symplectic curvature terms and symplectic covariant derivatives or either a membrane with a topologically nontrivial Maxwell contributions. This result we consider can be of interest for future phenomenological considerations. 

Geometrically the M2-brane on $M_9\times T^2$ with constant fluxes possesses three different structures of twisted torus $\mathbb{T}^3_W$ associated to the relation between the different fiber bundles over the worldvolume base manifold: one is associated to the nontrivial $U(1)$ bundle over the base, another is associated to the 2-torus target space over a homological one-cycle of the base with monodromy in $SL(2,Z)$ and a third one is associated to the torus on the target with a nontrivial flux.

We can define a Twisted torus bundle as a torus bundle with a $U(1)$ connection fibered over the base manifold $T^2_{U(1)}\to E\to \Sigma$. The M2-brane is consistently fiber over it. 
It allows to define the  monodromy of the bundle in $SL(2,Z)$. Compactification on  twisted torus have been shown to be related to Sherk-Schwarz reduction, metric fluxes and consequently with gauged supergravities at low energies. 
 We provide a concrete realization of this idea from M-theory, considering the Supermembrane theory on a twisted 3-torus  as related to 9D type II gauged supergravity. This construction clarifies previous results obtained in \cite{mpgm3} showing explicitly the relation between the constant fluxes, the central charge condition and the monodromy. It implies the correspondence between the classification of M2-brane inequivalent classes of torus bundles with monodromies in $SL(2,Z)$ and type II gauged supergravities in 9D.

 The fact that the non connected identity  symplectomorphisms (and not only the orientation preserved diffeomorphims) are the symmetries involved in the definition of the monodromy of the bundle (and respectively in  the classification of inequivalent M2-brane fiber bundles)    will have an impact in the classification of the $(p,q)$ strings that admit an M2-brane origin \cite{mpgm9}.

\section{Acknowledgements} 
The authors would like to thank to M. Asorey,  A. Guarino, F. Marchesano, P. Meessen, A. Uranga and A. Vi\~{n}a, for  helpful comments and discussions at different stages of this paper.  C.L.H want to also thanks to B. Fiol from Physics Department at  Barcelona U. for kind hospitality during the realization of part of
this work. M.P.G.M.  and P.L. also thank for  the to Theoretical Physics Department at Zaragoza U., for kind support and hospitality at initial  stages of this research. M.P.G.M also wants to thanks to IFT (CSIC-UAM), Madrid for her research stay -funded by MINEDUC-UA Project code ANT 1856, Antofagasta, Chile - where final part of this research was done. A.R. and M.P.G.M. are partially supported by Projects Fondecyt 1161192 (Chile), C.L.H, P.L. and J.M.P.  are supported by the Project ANT1955, ANT1756, ANT1855 and ANT1856 of the U. Antofagasta. P.L want to  thanks to CONICYT PFCHA/DOCTORADO BECAS CHILE/2019-21190517 and C.L.H thanks to CONICYT PFCHA/DOCTORADO BECAS CHILE/2019-21190263. The authors M.P.G.M., J.M.P., C.L.H, and P. L also thank to Semillero funding project SEM18-02 from U. Antofagasta, and to the international ICTP Network  NT08 for kind support.  

%
%
%





\end{document}